\documentclass[runningheads]{llncs}

\usepackage{tcolorbox}
\usepackage{algpseudocode} 
\usepackage{float}         
\usepackage{algorithm}
\usepackage{amsmath}
\usepackage{booktabs,multirow}
\usepackage{listings}
\usepackage{xcolor}
\usepackage{subcaption}
\usepackage{stmaryrd}
\usepackage[T1]{fontenc}

\definecolor{pygreen}{RGB}{0,120,0}
\definecolor{pyblue}{RGB}{0,0,180}
\definecolor{pyorange}{RGB}{180,90,0}
\definecolor{lightgray}{RGB}{248,248,248}
\definecolor{framegray}{RGB}{130,130,130}

\usepackage{marvosym}
\usepackage{orcidlink}
\usepackage{graphicx}
\begin{document}

\title{Prompt Optimization for LLM Code Generation via Reinforcement Learning}

%
%

\author{Ali Mohammadi Esfahani\inst{1}\textsuperscript{(\Letter)}\orcidlink{0009-0007-7770-2392} \and
Nafiseh Kahani\inst{1}\orcidlink{0000-0002-9322-0699} \and
Samuel A.Ajila\inst{1}\orcidlink{0000-0001-8824-1922} }
\authorrunning{A. Mohammadi Esfahani et al.}
%
\institute{Carleton University, Ottawa, ON, Canada \email{alimohammadiesfahani@cmail.carleton.ca}
}
\maketitle              

\begin{abstract}
Large Language Models (LLMs) can generate code from natural language, but their performance is highly sensitive to prompt formulation. We propose a reinforcement-learning-based framework that models prompt refinement as a sequential decision-making problem. A Proximal Policy Optimization (PPO) agent iteratively improves prompts using a hybrid action space that combines {direct generation,} genetic lexical mutation and semantic rewriting, guided by shaped rewards derived from unit-test feedback. We evaluate the framework on MBPP+, HumanEval+, and APPS using CodeT5+, CodeLLaMA, and DeepSeek-Coder as frozen code generators. On the 500-task MBPP+ test set, the PPO agent achieves {strict Pass@1 scores of 57.58\%, 64.80\%, and 85.50\%, respectively, outperforming EPiC, Reflexion, and Random-Hybrid. SoftPass@1 reaches 67.90\%, 73.10\%, and 88.20\%, respectively.} Similar improvements are observed on HumanEval+ and APPS across all backbone models. The results demonstrate that reinforcement learning with shaped test-driven rewards improves functional correctness in LLM-based code generation.

\keywords{Reinforcement Learning  \and Prompt Engineering \and Code Generation \and Large Language Models \and Genetic Algorithm.}
\end{abstract}

\section{Introduction}
\label{sec:Introduction}

Large Language Models (LLMs) have improved many automated software engineering tasks, particularly code generation from natural language descriptions \cite{r2,r3,r4,r5,r6,r7,r8,r9,r74}. However, generated code still frequently contains functional and logical defects, which limits its practical use. Recent studies have highlighted that even advanced LLMs exhibit relatively low exact-match accuracy, with critical defects frequently introduced into the generated code, requiring manual debugging and correction \cite{r62,r10}.

In a recent study, defects in LLM-generated code were investigated and categorized, with prevalent issues such as functionality, algorithmic, and logic defects being identified, emphasizing the necessity for targeted improvements \cite{r62}. The authors also demonstrated that standard prompt engineering techniques, such as Chain-of-Thought and Structured Chain-of-Thought prompting, can mitigate common defects by providing clearer instructions and context to the model \cite{r62}. However, these static prompting methods lack adaptability and often fall short when iterative refinement based on intermediate feedback is required.

To address these limitations, several adaptive approaches have recently been proposed, leveraging evolutionary algorithms or reinforcement learning (RL) frameworks for prompt optimization. Techniques such as EPiC \cite{r48} employ lightweight evolutionary search algorithms to iteratively refine prompts based on the performance of generated code evaluated against test cases. Similarly,  RL is used in Reflexion \cite{r63}, where language agents iteratively reflect on task feedback, incorporating self-generated linguistic insights into future prompt formulations. In addition, frameworks such as the Large Language Model Debugger (LDB) \cite{r64} enhance LLM-generated code by integrating runtime execution insights to iteratively pinpoint and correct defects.

Despite the effectiveness of these approaches, lexical or semantic prompt modifications are typically treated in isolation, and multiple dimensions of prompt transformations are not cohesively integrated. Furthermore, current strategies mainly rely on binary correctness signals, often overlooking the rich information contained in partial correctness or detailed feedback. {The contribution lies in coordinating reinforcement learning, genetic prompt mutation, and semantic rewriting through a learned sequential policy for executable code generation.} To fill these gaps, we propose an RL-driven prompt optimization approach specifically tailored for functional code generation tasks. This paper investigates whether RL can improve prompt optimization for code generation by adaptively combining lexical mutation and semantic rewriting. The proposed framework learns to select between lexical mutation and semantic rewriting based on intermediate test feedback. Prompt optimization is modeled as a sequential decision-making problem, where shaped rewards and {SoftPass@1 is used as an auxiliary learning-sensitive metric to} leverage partial correctness signals. The main contributions of this work are as follows.
\vspace{-0.08cm}
\begin{itemize}
    \item We formulate prompt optimization for code generation as a sequential decision-making problem and implement it as a PPO-based RL environment.
    
    \item We define a hybrid action space that combines direct generation, lexical mutation, and semantic rewriting for adaptive multi-step prompt refinement.
    
    \item We introduce a shaped reward function that uses partial test case correctness to provide denser learning signals than binary success alone.
    
    \item We evaluate the framework on MBPP+, HumanEval+, and APPS using CodeT5+, CodeLLaMA, and DeepSeek-Coder, and compare it against direct generation, EPiC, Reflexion, {and a Random-Hybrid baseline that uses the same action space without policy learning.}
\end{itemize}

The remainder of the paper is organized as follows. Section~\ref{sec:background} reviews related work. Section~\ref{sec:approach} presents the proposed framework. Section~\ref{sec:setup} describes the experimental setup and results. Finally, Section~\ref{sec:Conclusion} concludes the paper.

\section{Related work}
\label{sec:background}

Recent advances in LLMs have stimulated extensive research on prompt engineering and optimization for tasks such as code generation, instruction following, and program synthesis. Early approaches relied on manual prompt design, where human-crafted instructions were iteratively refined to elicit accurate model responses \cite{r27}, \cite{r5}. While effective in constrained settings, handcrafted prompts often lack generalization, motivating automated prompt optimization methods \cite{r13}, \cite{r51}, \cite{r50}. These approaches are broadly categorized into evolutionary search and RL frameworks.

Evolutionary methods treat prompt optimization as population-based search. EPiC \cite{r48} applies lexical mutations (insertion, deletion, replacement) guided by test-case correctness as a fitness signal. EvoPrompt \cite{r49} incorporates LLMs into mutation and crossover operations to produce semantically coherent variants, while PromptBreeder \cite{r50} introduces self-referential co-evolution of prompts and mutation operators. Although effective for exploring lexical diversity, evolutionary approaches generally rely on predefined operators and lack adaptive learning for multi-step refinement.

RL-based approaches formulate prompt optimization as sequential decision-making. RLPrompt \cite{r54} applies token-level edits using correctness-based rewards. PRewrite \cite{r52} extends the action space to include semantic rewriting via LLM-based paraphrasing. StablePrompt \cite{r56} improves training stability through Adaptive Proximal Policy Optimization, and PRL \cite{r61} explores RL-based prompt generation for few-shot tasks. However, most RL methods focus on narrow edit operations or fixed rewriting strategies and are typically evaluated on general language tasks rather than executable code generation with strict functional correctness constraints.

Existing methods either rely on lexical search, fixed semantic rewriting, or sparse correctness signals. Few combine heterogeneous prompt transformations within a learned sequential policy, and even fewer evaluate them on executable code generation with unit-test-driven rewards.
This work addresses those gaps by integrating lexical mutation and semantic rewriting in a unified PPO framework with shaped rewards based on partial correctness under unit-test evaluation. 

\section{Prompt Optimization  via Reinforcement Learning}
\label{sec:approach}

\begin{figure*}[t]
  \centering
  \includegraphics[width=\textwidth]{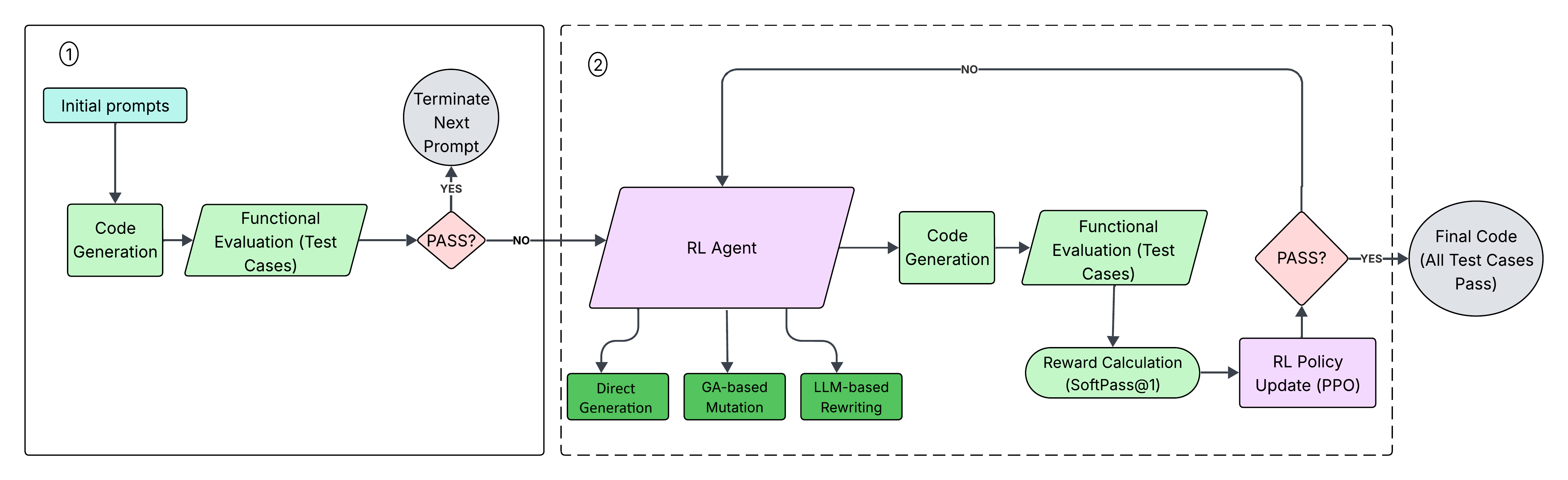}
  \caption{Workflow of the RL--based prompt optimization framework}
  \label{fig:workflow}
\end{figure*}


Prompt optimization is modeled as a sequential decision-making problem within an RL framework, where an agent iteratively refines natural language prompts in order to improve the functional correctness of code generated by an LLM.

In this setting, each programming task corresponds to an \textit{episode} of interaction with the environment. The \textit{state} represents the current prompt, encoded as a semantic embedding. The \textit{action space} consists of prompt transformation strategies that modify or reuse the current prompt. After a transformation is applied, the updated prompt is passed to a frozen code generation model to produce executable code. The generated program is evaluated against test cases, and the resulting correctness signal defines the \textit{reward}. The objective of the agent is to learn a policy that selects prompt transformations maximizing functional correctness across refinement steps.

The agent is trained using Proximal Policy Optimization \cite{r47}, selected for its stability and effectiveness in LLM-interactive environments. PPO is widely used in alignment and instruction tuning due to its controlled policy updates, computational efficiency, and robustness to noisy rewards \cite{r67}. The action space includes three actions: (1) Direct Generation, which preserves the current prompt, (2) genetic algorithm–based lexical mutation inspired by EPiC \cite{r48}, and (3) semantic rewriting inspired by Reflexion \cite{r63}.

During each episode, the environment operates in an iterative loop: the agent observes the current prompt representation, samples a prompt transformation from its policy, generates code using the updated prompt, evaluates the output against test cases, receives a reward based on functional correctness, and updates its policy accordingly. Unit tests are required during training to compute the reward signal and during evaluation to assess correctness, but they are not needed when the trained policy is applied for prompt refinement at inference time. Figure~\ref{fig:workflow} illustrates this test-driven prompt refinement process.\setlength{\parindent}{0pt}

\subsection{Environment Design}
\label{sec:Environment}

Prompt optimization is implemented within a custom Gymnasium-compatible RL environment. Each episode corresponds to a programming problem sampled from the training splits of MBPP+ \cite{r70}, HumanEval+ \cite{r70}, and APPS \cite{r71}, covering a range of task difficulties from simple algorithms to more complex multi-step reasoning tasks. An episode continues until either a fully correct program is generated or a predefined maximum number of refinement steps is reached.

At time step $t$, the agent observes the current prompt $p_t$. The prompt is embedded into a 384-dimensional vector $s_t$ using the MiniLM sentence transformer~\cite{r46}. This embedding serves as the state representation provided to the policy network and ensures a unified semantic representation across datasets. {This prompt-only state isolates prompt-level optimization, but it does not encode generated code, failed assertions, runtime exceptions, or tracebacks. Thus, the framework is best viewed as prompt-level refinement rather than traceback-guided debugging.}

At each step, the agent selects a prompt transformation, generating a modified prompt that is passed to the code generation model. The generated program at step $t$ of episode $i$ is denoted by $f_{i,t}$. The resulting function is evaluated against a set of test cases $\{(x_{i,k}, y_{i,k})\}_{k=1}^{K_i}$,  where $K_i$ denotes the number of unit tests associated with task $i$. For each test case, correctness is recorded as

\begin{equation}
c_{i,t,k} = \llbracket f_{i,t}(x_{i,k}) = y_{i,k} \rrbracket
\end{equation}

\noindent where $\llbracket \cdot \rrbracket$ denotes the indicator function.

The pass ratio for step $t$ in episode $i$ is then computed as

\begin{equation}
\rho_{i,t} = \frac{1}{K_i} \sum_{k=1}^{K_i} c_{i,t,k}.
\end{equation}

This pass ratio measures the proportion of test cases passed by the generated program and provides an intermediate signal of functional correctness. It forms the basis of the shaped reward used to guide policy learning.

\subsection{Action Space and Prompt Transformation Strategies}
\label{sec:action}

The agent operates in a discrete action space in which each action corresponds to a prompt transformation strategy. The proposed framework defines three actions. Direct generation is a no-edit control action, while lexical mutation and semantic rewriting are the two active refinement operators. The a priori motivation for combining the latter two is that they modify prompts along complementary dimensions: lexical mutation explores local surface-level variants, whereas semantic rewriting performs higher-level clarification and restructuring.

\textbf{Direct Generation} generates code using the unchanged prompt. Although no modification is applied, this action serves as a no-edit option that allows the policy to exploit decoding stochasticity and preserve prompts that are already close to optimal. Because autoregressive language models can produce different outputs from the same prompt under stochastic decoding, this action still supports useful exploration \cite{r66}.

\textbf{Lexical Mutation} applies a genetic algorithm inspired by EPiC \cite{r48}. The prompt is tokenized to form an initial population of candidate variants. Successive generations apply crossover, index shuffling, and mutation (with probability 0.2). Candidate prompts are evaluated based on the number of passed test cases, and tournament selection retains stronger variants for the next generation.

\textbf{Semantic Rewriting} performs higher-level reformulation using a pretrained model \cite{r58}, following the self-reflective paradigm of Reflexion \cite{r63}. A fixed meta-prompt guides structured rewriting of the original prompt, while regex-based filtering enforces basic structural constraints and ensures that task-relevant keywords are preserved.

At time step $t$, the agent samples an action $a_{i,t}$ from a stochastic policy $\pi_\theta$ conditioned on the current state $s_{i,t}$:

\begin{equation}
a_{i,t} \sim \pi_\theta(s_{i,t})
\end{equation}

{Here, $\pi_\theta$ outputs a probability distribution over the three available actions. For example, depending on the prompt embedding, the policy may assign higher probability to preserving the prompt, applying a lexical mutation, or invoking semantic rewriting.}

The selected action is applied via transformation operator $\mathcal{T}$ to obtain the updated prompt:

\begin{equation}
p_{i,t+1} = \mathcal{T}(p_{i,t}, a_{i,t})
\end{equation}

The operator $\mathcal{T}$ denotes the action-specific prompt update: it returns the unchanged prompt for direct generation, a locally mutated prompt for lexical mutation, or a semantically rewritten prompt for semantic rewriting.

The transformed prompt is then used for code generation, and the policy learns to sequence actions based on functional correctness feedback.

\subsection{Code Generation and Reward Structure}
\label{sec:Code}

After transformation, the prompt is passed to a code generation model to produce a candidate function. Each task includes test cases $\{(x_{i,k}, y_{i,k})\}_{k=1}^{K_i}$, and correctness is determined by executing the generated code against these tests. Execution is performed in a sandboxed environment adapted from Reflexion \cite{r63}, which safely captures syntax errors, runtime exceptions, and incorrect outputs. The outcome is used to compute a shaped reward based on the pass ratio $\rho_{i,t}$ (Section~\ref{sec:Environment}):

\begin{equation}
R_{i,t} =
\begin{cases}
1.0 & \text{if } \rho_{i,t} = 1.0 \\
\rho_{i,t} & \text{if } 0 < \rho_{i,t} < 1.0 \\
-1 & \text{if } \rho_{i,t} = 0 \\
-2 & \text{if execution fails}
\end{cases}
\end{equation}

This design rewards both full and partial correctness, penalizes complete functional failure, and strongly discourages execution errors. {Partial pass ratios are used only as intermediate learning signals for policy optimization; they are not treated as final functional success. Final task success is evaluated separately using strict Pass@1, which requires all tests to pass.} A stronger penalty is assigned to execution failure because syntactically invalid or non-executable outputs provide no usable functional signal for learning. In practice, execution failures receive a penalty of $-2$, which discourages the agent from producing syntactically invalid code while still allowing exploration of alternative prompt transformations. The cumulative episode reward is defined as
\begin{equation}
R^{\text{episode}}_i = \sum_{t=1}^{T_i} R_{i,t}.
\end{equation}

This reward signal encourages the agent to discover sequences of prompt transformations that progressively improve functional correctness. {The reward is therefore intended to guide search toward fully correct programs, not to accept partially correct programs as usable outputs.}

\subsection{Implementation}
Algorithm~\ref{alg:training} summarizes the training procedure of the RL-based prompt optimization framework. Each episode begins by selecting a programming task, which defines the environment instance. The initial prompt is embedded using MiniLM (Section~\ref{sec:Environment}) and fed to the policy network, providing a consistent semantic representation across tasks.

At each step, the agent selects an action as defined in Section~\ref{sec:action}. The transformed prompt is passed to a code generation model to produce executable Python code. The generated code is executed in a sandbox and evaluated against predefined test cases. A shaped reward is computed based on the proportion of passed tests (Section~\ref{sec:Code}), rewarding partial correctness and penalizing failures.

Each transition is stored for mini-batch updates, and at the end of the episode the policy is updated using PPO~\cite{r47}, ensuring stable and sample-efficient learning through clipped objectives.
{PPO policies were trained independently for each dataset--backbone configuration, using the corresponding training split and frozen code generator for that setting.}

\begin{figure}[t]
\centering

\begin{minipage}[t]{0.48\textwidth}
\begin{algorithm}[H]
\caption{Reinforcement Learning Training Loop}
\label{alg:training}
\begin{algorithmic}[1]
\Require Dataset $D$, PPO agent $\pi_\theta$, environment $E$, episodes $N$
\For{$i \gets 1$ to $N$}
  \State Sample task $(p_i, T_i)$ from $D$
  \State $s_{i,1} \gets \text{Embed}(p_i)$
  \For{$t \gets 1$ to $T_i$}
    \State $a_{i,t} \sim \pi_\theta(s_{i,t})$
    \State $p_{i,t+1} \gets T(p_{i,t}, a_{i,t})$ \Comment{transform prompt}
    \State $f_{i,t} \gets \text{LLM}(p_{i,t+1})$ \Comment{frozen code generator}
    \State Evaluate $f_{i,t}$ on $T_i$ to compute $\rho_{i,t}$
    \State Compute reward $R_{i,t}$ and store transition in $B_i$
    \State $s_{i,t+1} \gets \text{Embed}(p_{i,t+1})$
    \If{$\rho_{i,t} = 1.0$} \State \textbf{break} \EndIf
  \EndFor
  \State Update policy $\theta \gets \text{PPOUpdate}(\theta, B_i)$
\EndFor
\end{algorithmic}
\end{algorithm}
\end{minipage}
\hfill
\begin{minipage}[t]{0.48\textwidth}
\begin{algorithm}[H]
\caption{Evaluation Protocol for Prompt Optimization}
\label{alg:evaluation}
\begin{algorithmic}[1]
\Require Trained policy $\pi_\theta$, dataset $D_{\text{test}}$, step cap $T$
\State Initialize $C_{\text{strict}} \gets 0$, $S_{\text{soft}} \gets 0$
\For{each task $(p_i, T_i)$ in $D_{\text{test}}$}
  \State $s_{i,1} \gets \text{Embed}(p_i)$,\quad $P_i \gets 1$,\quad $\zeta_i \gets 0$
  \For{$t \gets 1$ to $T$}
    \State $a_{i,t} \sim \pi_\theta(s_{i,t})$
    \State $p_{i,t+1} \gets T(p_{i,t}, a_{i,t})$ \Comment{transform prompt}
    \State $f_{i,t} \gets \text{LLM}(p_{i,t+1})$
    \State Evaluate $f_{i,t}$ to compute $\rho_{i,t}$
    \State $P_i \gets P_i \cdot (1 - \rho_{i,t})$ \Comment{accumulate soft success}
    \If{$\rho_{i,t} = 1.0$}
      \State $\zeta_i \gets 1$ \Comment{mark strict success}
      \State \textbf{break}
    \EndIf
    \State $s_{i,t+1} \gets \text{Embed}(p_{i,t+1})$
  \EndFor
  \State $C_{\text{strict}} \gets C_{\text{strict}} + \zeta_i$
  \State $S_{\text{soft}} \gets S_{\text{soft}} + (1 - P_i)$
\EndFor
\State Compute metrics: $Pass@1_{\text{strict}} \gets \frac{C_{\text{strict}}}{|D_{\text{test}}|}$, \quad $Pass@1_{\text{soft}} \gets \frac{S_{\text{soft}}}{|D_{\text{test}}|}$
\end{algorithmic}
\end{algorithm}
\end{minipage}

\caption{(Left) PPO-based training loop; (Right) Evaluation protocol. Both algorithms are model-agnostic; the code generator is denoted by \textit{LLM}.}
\end{figure}

\section{Results and Analysis}
\label{sec:setup}
This section presents the research questions, experimental setup, and results. The results are organized to answer the following research questions:

\textbf{RQ1:} How do individual prompt transformation strategies perform in isolation when applied once without iterative feedback?

\textbf{RQ2:} How does the use of shaped reward functions and step-wise feedback influence the learning dynamics and performance of the RL agent?

\textbf{RQ3:} How effectively does multi-step RL improve prompt optimization compared to existing iterative strategies?

\subsection{Experimental Setup}

This section describes the experimental setup, including benchmarks, metrics, and baseline comparisons.

\paragraph{\textbf{Evaluation Metrics.}}

To evaluate the trained agent, we report strict Pass@1 and SoftPass@1. {Strict Pass@1 is treated as the primary measure of functional correctness because it counts a task as successful only when the generated solution passes all available tests. SoftPass@1 is used only as an auxiliary diagnostic metric for analyzing intermediate progress during multi-step refinement.}

For task $i$, let $\rho_{i,t}$ denote the fraction of test cases passed at refinement step $t$. Strict Pass@1 is computed as

\begin{equation}
\text{Pass@1}^{\text{strict}} =
\frac{1}{N}
\sum_{i=1}^{N}
\llbracket \exists t \in \{1,\dots,T_i\}: \rho_{i,t}=1.0 \rrbracket .
\end{equation}

This metric follows the standard interpretation of functional correctness: {a partially correct program is not considered functionally correct or deployable.} To capture partial progress during iterative refinement, SoftPass@1 is computed as

\begin{equation}
\text{Pass@1}^{\text{soft}} =
\frac{1}{N}
\sum_{i=1}^{N}
\left(1-\prod_{t=1}^{T_i}(1-\rho_{i,t})\right).
\end{equation}

{SoftPass@1 indicates whether the agent moves closer to full correctness across refinement steps. It should not be interpreted as evidence that partially correct code is practically usable.} The evaluation protocol used to compute these metrics is detailed in Algorithm~\ref{alg:evaluation}.

\paragraph{\textbf{Datasets.}}
Experiments were conducted on MBPP+~\cite{r70}, HumanEval+~\cite{r70}, and APPS~\cite{r71}. MBPP+ (974 tasks) was split into 374 training and 500 testing tasks. The remaining tasks were used for validation and excluded from training. For HumanEval+ and APPS, an 80\%/20\% train--test split was used.

\paragraph{\textbf{Model Selection.}}
Following the survey of LLMs for software engineering by Hou et al.~\cite{r18}, model selection was guided by explicit criteria. Only open-source models were considered to ensure reproducibility, excluding proprietary systems such as GPT-4~\cite{r1}, AlphaCode~\cite{r27}, PaLM-Coder~\cite{r11}, and Codex~\cite{r9}. Models not explicitly pre-trained on large-scale code corpora, including T5~\cite{r33}, GPT-Neo~\cite{r7}, and GPT-J~\cite{r41}, were also excluded. Older models superseded by updated versions, such as CodeT5~\cite{r43}, were not selected. Based on these criteria, CodeT5+~\cite{r17}, CodeLLaMA~\cite{r69}, and DeepSeek-Coder~\cite{r72} were chosen as code generation backbones.
\paragraph{\textbf{Baselines.}}
In addition to direct generation, EPiC, and Reflexion, we include Random-Hybrid as a non-learning ablation baseline. Random-Hybrid uses the same three actions as the proposed PPO framework. However, instead of learning a policy, it samples one action uniformly at random at each refinement step. We run Random-Hybrid for 10 refinement steps per task, using the same evaluation protocol as the PPO setting. This baseline isolates whether the performance gains come merely from access to the hybrid action space or from learned action scheduling.

\subsection{Experimental Results}

The proposed RL-based prompt optimization framework was evaluated on MBPP+, HumanEval+, and APPS using held-out test sets to assess its effectiveness across benchmarks of varying difficulty. The following subsections analyze the results with respect to the research questions, comparing static, heuristic, and learning-based approaches across benchmarks of increasing difficulty.

\begin{table*}[ht]
\centering
\caption{Comparison of Pass@1 and SoftPass@1 across datasets and prompt optimization actions. {Semantic rewriting generally provides the strongest single-step improvement, while SoftPass@1 highlights partial functional progress beyond strict Pass@1.}}
\label{tab:multi_dataset_results}

\resizebox{\textwidth}{!}{
\begin{tabular}{@{}llcccccc@{}}
\toprule
\textbf{Dataset} & \textbf{Method} 
& \multicolumn{2}{c}{\textbf{CodeT5+}} 
& \multicolumn{2}{c}{\textbf{CodeLLaMA}} 
& \multicolumn{2}{c}{\textbf{DeepSeek-Coder}} \\
\cmidrule(lr){3-4} \cmidrule(lr){5-6} \cmidrule(lr){7-8}
 & 
& \textbf{Pass@1} & \textbf{SoftPass@1}
& \textbf{Pass@1} & \textbf{SoftPass@1}
& \textbf{Pass@1} & \textbf{SoftPass@1} \\
\midrule

\multirow{3}{*}{\textbf{MBPP+}}
& Direct Generation  &12.84\% & 22.80\% & 41.91\% & 48.82\% & 76.10\% & 77.86\% \\
& Genetic Mutation   & 20.23\% & 33.40\% & 43.01\% & 51.70\% & 78.24\% & 78.79\% \\
& Semantic Rewriting & 23.73\% & 36.90\% & 46.20\% & 55.20\% & 78.49\% & 79.18\% \\
\midrule

\multirow{3}{*}{\textbf{HumanEval+}}
& Direct Generation  & 15.10\% & 19.61\% & 33.02\% & 35.61\% & 89.90\% & 90.01\% \\
& Genetic Mutation   & 21.49\% & 30.75\% & 34.63\% & 40.81\% & 89.24\% & 89.49\% \\
& Semantic Rewriting & 25.59\% & 37.12\% & 32.33\% & 38.91\% & 89.49\% & 89.94\% \\
\midrule

\multirow{3}{*}{\textbf{APPS}}
& Direct Generation  & 8.20\% & 14.60\% & 14.75\% & 22.10\% & 18.30\% & 27.40\% \\
& Genetic Mutation   & 10.95\% & 19.80\% & 17.90\% & 27.60\% & 21.75\% & 32.50\% \\
& Semantic Rewriting & 13.40\% & 23.70\% & 20.85\% & 31.20\% & 24.60\% & 36.80\% \\
\bottomrule
\end{tabular}
}
\end{table*}

\vspace{0.1cm}\noindent\textbf{RQ1: Performance of Prompt Transformations}

\noindent We evaluate three single-step prompt transformation strategies without iterative refinement. Each benchmark task is processed once and evaluated using strict Pass@1. Quantitative results are provided in Table~\ref{tab:multi_dataset_results}.

Across most datasets and backbone models, semantic rewriting yields the strongest single-step improvements, followed by lexical mutation, while direct prompting performs the weakest. These gains suggest that reducing ambiguity and restructuring instructions increases the likelihood of generating functionally correct programs. {Although lexical mutation is less effective than semantic rewriting in a single-step setting, it still improves over direct generation and provides useful local exploration. This suggests that mutation and rewriting play complementary roles: rewriting introduces larger semantic clarification, whereas mutation explores smaller lexical variants around a promising prompt.}

Performance decreases on APPS due to its higher algorithmic complexity, yet the relative ordering of strategies remains largely consistent. This consistency indicates that structured prompt clarification provides a robust one-shot enhancement mechanism across difficulty levels.

Overall, while single-step transformations improve performance, their impact remains limited without iterative feedback. These findings motivate the need for adaptive, multi-step refinement, as explored in the subsequent RL framework. {The Random-Hybrid and PPO comparisons in RQ3 further test whether simply having both transformation types is sufficient, or whether learned action scheduling is needed to use them effectively.}

\begin{figure*}[ht]
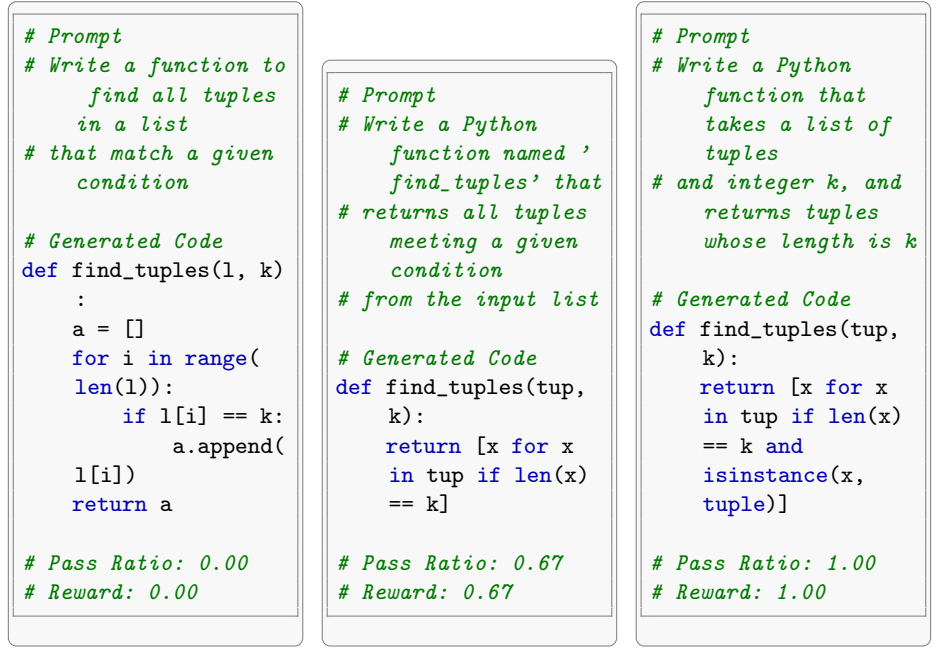

\centering

\begin{subfigure}[t]{0.32\textwidth}
\begin{tcolorbox}[
  colback=lightgray,
  colframe=framegray,
  boxrule=0.4pt,
  left=2pt,right=2pt,top=2pt,bottom=2pt
]
\begin{lstlisting}
# Prompt
# Write a function to find all tuples in a list
# that match a given condition

# Generated Code
def find_tuples(l, k):
    a = []
    for i in range(len(l)):
        if l[i] == k:
            a.append(l[i])
    return a

# Pass Ratio: 0.00
# Reward: 0.00
\end{lstlisting}
\end{tcolorbox}
\caption{Original Prompt (Ambiguous, No Pass)}
\label{fig:rq3a}
\end{subfigure}
\hfill
\begin{subfigure}[t]{0.32\textwidth}
\begin{tcolorbox}[
  colback=lightgray,
  colframe=framegray,
  boxrule=0.4pt,
  left=2pt,right=2pt,top=2pt,bottom=2pt
]
\begin{lstlisting}
# Prompt
# Write a Python function named 'find_tuples' that
# returns all tuples meeting a given condition
# from the input list

# Generated Code
def find_tuples(tup, k):
    return [x for x in tup if len(x) == k]

# Pass Ratio: 0.67
# Reward: 0.67
\end{lstlisting}
\end{tcolorbox}
\caption{Partial Pass after Semantic Rewrite}
\label{fig:rq3b}
\end{subfigure}
\hfill
\begin{subfigure}[t]{0.32\textwidth}
\begin{tcolorbox}[
  colback=lightgray,
  colframe=framegray,
  boxrule=0.4pt,
  left=2pt,right=2pt,top=2pt,bottom=2pt
]
\begin{lstlisting}
# Prompt
# Write a Python function that takes a list of tuples
# and integer k, and returns tuples whose length is k

# Generated Code
def find_tuples(tup, k):
    return [x for x in tup if len(x) == k and isinstance(x, tuple)]

# Pass Ratio: 1.00
# Reward: 1.00
\end{lstlisting}
\end{tcolorbox}
\caption{Full Pass after Refinement}
\label{fig:rq3c}
\end{subfigure}

\caption{Illustrative progression from an ambiguous benchmark-style prompt to partial and full correctness, showing the corresponding generated code, pass ratios, and rewards.}
\label{fig:rq3_prompt_example}
\end{figure*}

\vspace{0.1cm}\noindent\textbf{RQ2. Shaped Rewards and Feedback}

\noindent To evaluate the impact of reward design, we compare a sparse binary reward with the proposed shaped reward under identical environments and action spaces. In the binary setting, the agent receives a reward of $1.0$ only if all test cases pass; otherwise, it receives zero. In contrast, the shaped reward assigns proportional credit based on the fraction of passed test cases at each step, with additional penalties for execution failures. This formulation provides continuous feedback rather than episodic success-only signals.

Empirically, reward shaping leads to consistently stronger performance across datasets and backbone models. On MBPP+, the shaped-reward agent achieves substantial improvements under both strict Pass@1 and SoftPass@1 (see Table~\ref{tab:rq3_rq4_combined}). In contrast, the binary-reward variant converges more slowly and exhibits higher instability during training, indicating that sparse success signals are insufficient for effective policy learning in multi-step prompt refinement.

Step-wise feedback plays a central role in this improvement. Because the shaped reward provides proportional credit for partial correctness, the agent can detect when a semantic rewrite or lexical mutation increases the number of passing test cases. This enables refinement of promising transformations rather than discarding them prematurely. Under the binary condition, partially correct outputs are treated as complete failures, preventing incremental learning and discouraging multi-step exploration.

The effect is further illustrated in Figure~\ref{fig:rq3_prompt_example}, {which shows an illustrative benchmark-style refinement trajectory} where intermediate prompt transformations progressively increase the pass ratio. The shaped reward reinforces these intermediate gains, allowing the policy to accumulate functional progress even when full correctness is not immediately achieved.

Reward shaping stabilizes learning, supports incremental credit assignment, and is therefore used in all subsequent experiments.

\begin{table*}[ht]
\centering
\caption{Comparison of PPO against EPiC, Reflexion, and Random-Hybrid across MBPP+, HumanEval+, and APPS. Bold values indicate the proposed PPO framework.}
\label{tab:rq3_rq4_combined}

\resizebox{\textwidth}{!}{
\begin{tabular}{@{}llcccccc@{}}
\toprule
\textbf{Dataset} & \textbf{Method} 
& \multicolumn{2}{c}{\textbf{CodeT5+}} 
& \multicolumn{2}{c}{\textbf{CodeLLaMA}} 
& \multicolumn{2}{c}{\textbf{DeepSeek-Coder}} \\
\cmidrule(lr){3-4} \cmidrule(lr){5-6} \cmidrule(lr){7-8}
 & 
& \textbf{Pass@1} & \textbf{SoftPass@1}
& \textbf{Pass@1} & \textbf{SoftPass@1}
& \textbf{Pass@1} & \textbf{SoftPass@1} \\
\midrule

\multirow{4}{*}{\textbf{MBPP+}}
& EPiC      
& 41.89\% & 54.20\% 
& 51.40\% & 61.50\% 
& 80.30\% & 82.10\% \\
& Reflexion 
& 41.63\% & 55.10\% 
& 52.70\% & 63.60\% 
& 81.60\% & 83.40\% \\
&  {Random-Hybrid}
&  {31.12\%} &  {44.10\%}
&  {49.08\%} &  {58.60\%}
&  {79.10\%} &  {79.90\%} \\
& PPO (Ours)
& \textbf{57.58\%} & \textbf{67.90\%}
& \textbf{64.80\%} & \textbf{73.10\%}
& \textbf{85.50\%} & \textbf{88.20\%} \\
\midrule

\multirow{4}{*}{\textbf{HumanEval+}}
& EPiC      
& 33.80\% & 46.90\% 
& 41.20\% & 52.60\% 
& 90.10\% & 91.80\% \\
& Reflexion 
& 35.40\% & 48.70\% 
& 43.00\% & 54.90\% 
& 91.10\% & 92.90\% \\
&  {Random-Hybrid}
&  {31.10\%} &  {42.60\%}
&  {38.10\%} &  {45.80\%}
&  {89.60\%} &  {90.10\%} \\
& PPO (Ours)
& \textbf{60.20\%} & \textbf{69.80\%}
& \textbf{70.60\%} & \textbf{77.90\%}
& \textbf{91.30\%} & \textbf{92.50\%} \\
\midrule

\multirow{4}{*}{\textbf{APPS}}
& EPiC      
& 16.40\% & 28.75\% 
& 24.80\% & 38.60\% 
& 29.90\% & 45.10\% \\
& Reflexion 
& 17.60\% & 30.10\% 
& 26.10\% & 40.20\% 
& 31.20\% & 47.30\% \\
&  {Random-Hybrid}
&  {13.90\%} &  {24.10\%}
&  {21.50\%} &  {31.90\%}
&  {24.70\%} &  {37.50\%} \\
& PPO (Ours)
& \textbf{20.75\%} & \textbf{35.90\%}
& \textbf{29.80\%} & \textbf{45.60\%}
& \textbf{34.95\%} & \textbf{52.40\%} \\
\bottomrule
\end{tabular}
}
\end{table*}

\vspace{0.1cm}\noindent\textbf{RQ3. Multi-Step RL Effectiveness}

\noindent We evaluate the proposed PPO-based RL framework against EPiC, Reflexion, and  {Random-Hybrid} across MBPP+, HumanEval+, and APPS. All backbone models remain frozen during evaluation, and performance is measured using both strict Pass@1 and SoftPass@1. Detailed results are reported in Table~\ref{tab:rq3_rq4_combined}.  {Random-Hybrid uses the same three-action space as PPO but selects actions uniformly at random for 10 refinement steps, isolating the effect of learned action scheduling.}

Across all datasets and backbone models, PPO outperforms EPiC, Reflexion, and  {Random-Hybrid}.  {Since Random-Hybrid has access to the same transformations as PPO, this result shows that the gains are not explained by the hybrid action space alone.} On MBPP+, PPO improves strict Pass@1 over Random-Hybrid by  {26.46 percentage points for CodeT5+, 15.72 points for CodeLLaMA, and 6.40 points for DeepSeek-Coder}. Similar trends are observed on HumanEval+ and APPS, although absolute performance decreases on APPS due to its higher algorithmic complexity and stricter evaluation protocol. These results indicate that learned multi-step coordination is more effective than mutation-driven, rewrite-driven, or randomly scheduled hybrid refinement.

SoftPass@1 follows the same trend. Beyond increasing strict task completion rates, PPO accumulates stronger intermediate functional progress, suggesting that the learned policy can preserve and build upon partial improvements during iterative prompt refinement.  {This is particularly important because PPO and Random-Hybrid share the same available transformations; the performance gap therefore reflects learning when to apply each transformation rather than merely combining them.}

To assess statistical significance, paired comparisons were conducted across tasks in the evaluation sets. For strict Pass@1, McNemar’s test was applied on the MBPP+ test set (500 tasks), yielding statistically significant improvements over both EPiC and Reflexion ($p < 0.001$) with medium effect sizes (Cohen’s $h$ approximately 0.27--0.35). For SoftPass@1, paired $t$-tests indicate highly significant gains ($p < 0.0001$) with medium-to-large effect sizes (Cohen’s $d$ approximately 0.60--0.75). Overall, the results suggest that PPO improves prompt optimization by adaptively sequencing lexical and semantic transformations based on feedback signals.

\subsection{Threats to Validity}
\label{sec:validity}

Several factors may affect the generality of the results. First, the evaluation relies on benchmark datasets and predefined test suites, which may not fully capture real-world programming complexity. Second, the code generation models remain frozen during training; therefore, improvements arise only from prompt optimization rather than model adaptation. Third, the quality of semantic rewriting depends on the rewriting language model and the structure of the meta-prompt. Finally, the RL state representation uses only prompt embeddings and does not explicitly incorporate previous actions or reward history.  {It also excludes generated code and execution tracebacks, limiting diagnostic refinement. In addition, semantic rewriting and PPO training increase computational cost because they require repeated rewriting, generation, execution, and reward computation.}

\section{Conclusion}
\label{sec:Conclusion}

This paper presented a reinforcement learning framework for multi-step prompt optimization in code generation. By modeling prompt refinement as a sequential decision-making process and combining lexical mutation with semantic rewriting, the proposed PPO-based agent consistently outperformed EPiC, Reflexion, and  {Random-Hybrid} across MBPP+, HumanEval+, and APPS.  {The Random-Hybrid comparison shows that the gains come not only from access to multiple transformation operators, but also from learned action scheduling.} These results highlight the value of adaptive, feedback-driven prompt optimization, where the agent dynamically coordinates semantic rewriting and lexical refinement using shaped rewards and intermediate correctness signals. Consistent gains across multiple backbone models indicate robustness to architectural variation and task difficulty. {The current framework remains limited by its prompt-only state representation, which does not explicitly encode generated code, failed tests, or execution tracebacks.} Future work may extend the framework to larger benchmarks, transfer or curriculum learning,  {traceback-aware state representations,} and practical development-tool integration.

\section*{Acknowledgements}
We thank the Natural Sciences and Engineering Research Council of Canada (NSERC) for supporting this work.

%
%
%


%
%
%
\bibliographystyle{splncs04}
\bibliography{mybibliography}
%




\end{document}